\documentclass[preprint,showpacs,preprintnumbers,amsmath,amssymb,nofootinbib,showkeys,aps]{revtex4}
\pdfoutput=1
\usepackage{graphicx}% Include figure files
\usepackage{dcolumn}% Align table columns on decimal point
\usepackage{bm}
\usepackage{epsfig}

\newcommand{\beq}{\begin{equation}}
\newcommand{\eeq}{\vspace{0cm} \end{equation}}
\newcommand{\beqq}{\setlength\arraycolsep{2pt}\begin{eqnarray}}
\newcommand{\eeqq}{\vspace{0cm} \end{eqnarray}}

\usepackage{hyperref}

\begin{document}
\title{Cosmic Expansion Driven by Gravitational Particle Production: Toward a Complete Cosmological Scenario}

\author{P. W. R. Lima$^{1}$\footnote{pablolima@usp.br}}
\author{J. A. S. Lima$^{1}$\footnote{jas.lima@iag.usp.br}}
\affiliation{$^{1}$Departamento de Astronomia, Universidade de S\~{a}o
Paulo \\ Rua do Mat\~ao, 1226 - 05508-900, S\~ao Paulo, SP, Brazil}

\keywords{Particle creation, Accelerating CCDM cosmology, Dark Sector Reduction}

\bigskip
\begin{abstract}
A dark-energy-free cosmological model ($\Omega_{DE} \equiv 0$) based on gravitationally induced adiabatic particle creation is proposed. The thermodynamics of particle production yields an effective negative pressure that drives both primordial inflation and late-time cosmic acceleration. The model, characterized by four components and two free parameters ($\alpha$, $\beta$), reproduces a $\Lambda$CDM-like expansion for suitable $\alpha$, while $\beta$ introduces small but testable deviations from the cosmic concordance model. Constraints from type Ia Supernovae (Pantheon+SH0ES) and $H(z)$ data indicate $\beta \simeq 0.13$, suggesting a mild departure from standard cosmology and possible relief of the $H_0$ and $S_8$ tensions. The resulting classical cosmology evolves smoothly between two extreme de Sitter phases, offering a singularity-free, unified scenario that beyond solving old cosmological puzzles opens a new perspective to handle the tensions  plaguing the current cosmic concordance model. 

\end{abstract}

\maketitle

\section{Introduction}
Although remarkably successful, simple, and quite predictive, the prevailing cosmological paradigm ($\Lambda$CDM supplemented by inflation) can be viewed as a patchwork of distinct components assembled to describe cosmic evolution across all times and length scales. Such disparate ingredients are invoked to connect, without discontinuity, the accelerating phases at the early and late times of the expanding universe \cite{Planck2018a,Planck2018b,Peri2022}. 

From a theoretical perspective, beyond the long-standing cosmological constant problem (CCP) and cosmic coincidence mystery (CCM), the concordance model also faces several interrelated conceptual challenges. These include the initial singularity, the graceful-exit problem in realistic inflationary scenarios, and trans-Planckian issues that remain unresolved from first principles \cite{LBS2013,P2013,RHBM2013,Vafa2020,Tomi2020,Dine2021}.

On the  observational front, $\Lambda$CDM is also increasingly challenged  by the persistent $H_0$ and $S_8$ tensions. The former reflects a significant and sustained discrepancy between local distance-ladder measurements of the Hubble constant $H_0$ and its early-universe inference from CMB data within the $\Lambda$CDM framework \cite{Verde2019,Riess2022,A2022}. The latter discrepancy arises because late-time large-scale-structure measurements, based on cosmic shear (e.g., weak lensing and galaxy clustering), favor lower values of $S_8 \equiv \sigma_8 \sqrt{\Omega_M/0.3}$ - where $\sigma_8$  is the current mass fluctuation in $8h^{-1}\,\mathrm{Mpc}$ scales - as compared with the values inferred from early universe probes \cite{DESY32023,Brout2022,Heymans2021,Alam2021}. Although unresolved systematic effects cannot be entirely excluded, the combined pattern of tensions has inspired  extensions to $\Lambda$CDM, such as dynamical dark energy, multi-interacting dark energy models, decaying-vacuum scenarios, or new physics affecting cosmic expansion and structure growth \cite{Diva2021, Lucca2021, PLS2024,SOT2010, CAPO2011,H2015, LBS2016,Lomb2019,Rafael2020,Masoori2024}. 

It is worth noticing that negative pressure is the essential ingredient that drives the accelerated expansion of the Universe and, in the standard picture, provides the main observational motivation for dark energy. However, such stresses also arise naturally in nonequilibrium systems, especially during phase transitions — for example, in an overheated van der Waals fluid — where negative pressure can become unavoidable. In cosmology, as first suggested  by Zel’dovich \cite{ZED70}, particle creation by the gravitational field of the expanding Universe, can phenomenologically be described by an effective negative pressure associated with bulk viscosity and entropy production.

%It is worth noticing should also be recalled that the presence of a negative pressure is the key ingredient required to accelerate the expansion of the Universe. In the standard view, it is the main observation directly justifying the dark energy, an ubiquitous cosmic dominant component with negative pressure.  However, such a stress also arises naturally in different physical contexts whenever a system departs from thermodynamic equilibrium. Usually, these nonequilibrium states are associated with phase transitions—for example, in an overheated van der Waals liquid—where the emergence of a negative pressure is not only possible, but inevitable in some cases. In cosmology, as first emphasized by Zeldovich \cite{ZED70}, the process of particle creation at the expense of the gravitational field can be phenomenologically described by the effective negative pressure of the bulk viscosity mechanism, which is accompanied by entropy production. 

Later, the  phenomenological macroscopic suggestion of Zeldovich was realized in a more consistent formulation by introducing a source term describing the particle production, as well as the back reaction in spacetime through a negative creation pressure \cite{PRI89,LCW91,CLW92,LG92}. This interpretation established a natural thermodynamic bridge between spacetime dynamics and the irreversible creation of matter. Microscopically, the influence of gravitationally induced matter creation (GIMC) process was first investigated by Parker, who employed the Bogoliubov mode-mixing technique within the framework of quantum field theory in curved spacetime \cite{P68,P69,P74}. This method was further explored and also extended by many authors (see \cite{D82,MW07,HU09} and Refs. therein). Although these models were rigorous and well-motivated, they were never fully realized  due to absence of a clear prescription for incorporating matter creation and its back reaction in the classical Einstein Field Equations (EFE). 

Interestingly, the growing tension between observation and theory within the $\Lambda$CDM paradigm has also prompted a closer scrutiny of the dark sector itself. Instead more elusive forms of dark energy, some authors have proposed a radical constraint by setting $\Omega_{DE} \equiv 0$.  Unlike dark energy, the basic idea here is that DM component is needed in practically all relevant scales (from galaxies to cosmology). By replacing an unknown dark energy component with a mechanism rooted in microscopic physics (continuous matter creation), the model transforms the question of cosmic acceleration from a problem of missing energy into one of fundamental physical process. 

Naturally, such an approach also opens a new conceptual path toward resolving the persistent challenges that confront not only $\Lambda$CDM but any cosmology whose acceleration at late times is powered by dark energy. Actually, during the past decade, motivated by models with matter creation that successfully mimic the $\Lambda$CDM dynamics in the absence of dark energy \cite{LJO10,LBC12,Pert1,Waga2,Pert2}, several authors have undertaken a more systematic investigation of cosmologies with matter creation, addressing a broad range of theoretical and observational objectives (see \cite{LGPB14,  LB2014,GCL2014,BL2015,Mukherjee2016,S2016,LS2016,SPan2016,jesus2017,LT2021,Cardenas2021,TL2023,EKO2024,Cardenas2025,LLJ2025}  for an incomplete list). 

It is also worth noticing that the irreversible matter creation process, in the unperturbed level, introduces two key modifications in comparison to $\Lambda$CDM equilibrium cosmology: (i) a balance equation for the particle number density, and (ii) a negative pressure term in the energy-momentum tensor, and entropy growth. These three quantities are closely linked through the second law of thermodynamics. The central idea of this framework is that matter creation, driven by the gravitational field, can only occur as an irreversible process constrained by the principles of non-equilibrium thermodynamics.

In this framework, we propose a flat cosmological model in which the present acceleration arises solely from the creation of cold particles—dark matter and baryons. The resulting classical cosmology is dynamically complete, evolving between two de Sitter phases. As shown below, the model fits Type Ia supernova data and predicts a transition redshift of order unity. In this extended CCDM scenario, the Hubble parameter need not be fine-tuned to solve the age problem, and cosmic acceleration naturally emerges even when matter creation is negligible during the radiation era and most of the matter-dominated epoch. Moreover, the coincidence problem of dark energy cosmologies is replaced by a gravitationally induced particle-creation process operating at low redshifts.

The present article is planned as follows. Section II provides a brief review of cosmic macroscopic particle production, as well as the fundamental dynamical and thermodynamical aspects of a flat FRW cosmology. In Section III, we introduce the basic phenomenological creation rates that yield a classical and complete cosmic description. Section IV examines in detail the distinct cosmological eras after Planck time, while Section V presents our statistical analysis focusing on the matter-dominated epoch. Finally, Section VI summarizes the main results and conclusions concerning the classically complete accelerating cosmological model without dark energy ($\Omega_{DE} \equiv 0$), driven solely by gravitationally induced particle production.

\section{Cosmic Dynamics and Thermodynamics with Particle Production}\label{PC}

In this section, we briefly review  the dynamics and thermodynamics of the universe, with particular emphasis on cosmological fluids endowed with ``adiabatic'' particle production.   

Initially, let us assume a homogeneous and isotropic universe described by the flat Friedmann-Lema\^itre-Robertson-Walker (FLRW) geometry:
\begin{equation}
ds^{2} = dt^{2} - a^{2}(t) \left(dx^{2} + dy^{2} + dz^{2}\right),
\label{ds2}
\end{equation}
where $a(t)$ is the scale factor. The Einstein Field Equations (\textbf{EFE}) with regards to the geometry described above, for a single fluid endowed with particle creation, can be written in natural units as (for multi-fluid cosmology with particle production see \cite{Pert2,TL2023})
\begin{align}
8\pi G \rho &= 3H^2,\label{EE1}\\
8\pi G (p + p_c) &= -2\dot{H} - 3H^{2},\label{EE2}
\end{align}
where an overdot means derivative with respect to comoving time, $G$ is the gravitational constant and $\rho$, $p$, $p_c$ and $H=\dot{a}/a$ are respectively the energy density, equilibrium pressure, creation pressure and the Hubble parameter. 

In the standard formulation, the thermodynamic description of a simple relativistic fluid is governed by the energy-conservation law (ECL), $u_{\mu} T^{\mu\nu}_{;\nu}= 0$, together with the conservation of particle number and entropy fluxes,   $N^{\mu}_{;\mu}=0$ and $S^{\mu}_{;\mu}=0$. However, in the more general picture where a mechanism of gravitationally induced particle production is present, the fluxes $N^{\mu}$ and $S^{\mu}$ no longer satisfy strict conservation laws and become balance equations \cite{LT2021,TL2023}

\begin{eqnarray}\label{TEQ}
u_{\mu} T^{\mu\nu}_{;\nu}&=& 0 \,\,\,\, \Longleftrightarrow \,\,\,\, \dot{\rho} + 3H(\rho + p + p_c)=0,   \label{n_eq} 
\\
N^{\mu}_{;\mu} &=& n\Gamma_N \,\,\,\, \Longleftrightarrow \,\,\,\,\ \dot n + 3nH = n\Gamma_N
 \Longleftrightarrow \frac{\dot N}{N} = \Gamma_N,  \label{n_gamma}
\\ 
S^{\mu}_{;\mu} &=& s\Gamma_S \,\,\,\, \Longleftrightarrow \,\,\,\,\ \dot s + 3sH = s\Gamma_S\,\, \Longleftrightarrow \frac{\dot S}{S} = \Gamma_S,   \label{s_gamma}
\end{eqnarray}
in which $n=Na^{-3}$ and $s=Sa^{-3}$ are the total number of particles and entropy in a commoving volume $V\propto a^{3}$, whereas $\Gamma_N$ and $\Gamma_S$ are the particle and entropy creation rates, respectively. 
In this context, let us investigate the ``adiabatic'' process, that is, the case in which the entropy per particle $\sigma=S/N$ is conserved ($\dot{\sigma}=0$). It follows from equations \ref{n_gamma} and \ref{s_gamma} that
\begin{equation}
   \dot \sigma =  \sigma \left(\frac{\dot S}{S}-\frac{\dot N}{N}\right) = 0,  
\end{equation}
which implies the relations:
\begin{equation}\label{sngamma}
    \frac{\dot S}{S} = \frac{\dot N}{N}  = \Gamma \geq 0,
\end{equation}
where $\Gamma=\Gamma_N =\Gamma_S$ and the inequality in the equation above is imposed by the second law of thermodynamics. The imposition $\Gamma\geq0$ ensures that gravitational effects can only create, never annihilate, particles and entropy in this scenario. Moreover, integration of equation (\ref{sngamma}) implies that $S=k_B N$, where $k_B$ is the Boltzmann constant.   

An expression for the particle creation pressure $p_c$ can be easily obtained from the previous equations and the well-known local Gibbs law: $nTd\sigma =  d\rho - (\rho + p)dn/n$, where T is the temperature. This law, when combined with equations (\ref{n_eq}) and (\ref{n_gamma}) in the case $\dot\sigma=0$, yields  \cite{LJO10,LBC12,Pert1,Pert2,LT2021,TL2023}
\begin{equation}\label{Pc1}
p_{c}=-(\rho + p)\, \frac{\Gamma}{3H}\, .
\end{equation}
Note that $p_c$ is always negative in the ``adiabatic'' scenario, provided that $\Gamma$, $p$, $\rho$ and $H$ are all positive define quantities. This is a reasonable statement when assuming the simplest Equation of State (EoS), without any exotic fluid (e.g. dark energy), that is 
\begin{equation}\label{EoS}
p=\omega\rho, \,\,\,\omega \geq 0.
\end{equation}

The combination of the equations obtained above, specially equations (\ref{n_eq}), (\ref{Pc1}) and (\ref{EoS}), allows us to write the differential equation that determines the evolution of the energy density as
\begin{equation}\label{rhodot}
\dot{\rho}+3H\rho(1+\omega)\left(1-\frac{\Gamma}{3H}\right)=0 \,.
\end{equation}
It is also convenient to rewrite the above differential equation for the evolution of $H$ directly from equation (\ref{EE1}). It reads:
\begin{equation}\label{Hdot}
\dot{H} +\frac{3(1+w)}{2}H^{2}\left(1-\frac{\Gamma}{3H}\right)=0 .
\end{equation}
Thus, in the absence of particle creation ($\Gamma = 0$), equation (\ref{rhodot} recovers the usual flat FLRW case. In addition, the extreme creation rate,  $\Gamma = 3H$, results in  $\dot{\rho}=\dot{H}=0$, that is, $\rho$ and $H$ constants, regardless of the value of $\omega$. In other words, ``adiabatic'' particle creation allows de Sitter cosmic expansions driven by regular matter-energy contents, such as radiation or even dust formed by cold dark matter and/or baryons \cite{PRI89,LCW91,CLW92,LG92,LJO10}. 

%, (\ref{EE2}), (\ref{Pc1}) and (\ref{EoS})

As will be shown next, an evolution from a primordial de Sitter inflationary phase ($H=H_I$) to a late time de Sitter stage ($H=H_F$) can rigorously be defined without dark energy. The resulting scenario is classically complete, by assuming that the primeval de Sitter state starts just after the quantum gravity regime. In this connection, as proposed long ago by Vilenkin \cite{V1985},  the classical Universe may emerge from ``nothing" as a de Sitter solution from an, expected but unknown quantum gravity regime. Here, it will be characterized by the energy scale $H_I$, limited by the reduced Planck mass $H_I \leq M_{Pl} = (8 \pi G)^{-\frac{1}{2}} \simeq 2.4\times 10^{18} GeV$, in agreement with the Friedmann equation ($\ref{EE1}$). 

%In Figure \textbf{1}, we  display the timeline with the key events for this complete classical evolution (from de Sitter to de Sitter) powered by the creation of matter. Now, let us discuss the most interesting details. 
\begin{figure}[!ht]
\centering
\includegraphics[scale=0.65]{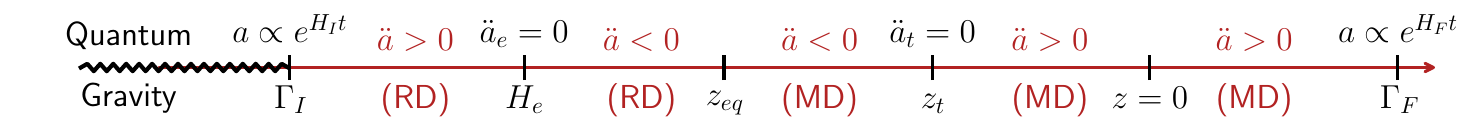}
\caption{Timeline  of the complete cosmological scenario from the initial Sitter ($H_I$)  to a final Sitter ($H_F$) stage (out of visual scale). Immediately after the unknown primordial quantum gravity state, ultra relativistic particles and radiation are created at the expenses of gravity (curvature) with the rate $\Gamma_I=3H_I$ and inflate the cosmos. Inflation ends ($\ddot a=0$) for $H=H_e$,  when the decelerating radiation dominated (RD) phase begins ($\ddot a<0$). This RD regime with creation ends at $z=z_{eq}$ marking the begin of the decelerating matter dominated (MD) phase with creation. At $z_{t}$, the creation pressure accelerates the cosmic expansion once more, which is observed today ($z=0$), until it finally induces a final de Sitter-like stage powered by $\Gamma_F=3H_F$ in the very distant future. Figure adapted from \cite{PLS2024}.}
\label{timeline}
\end{figure}

In Figure \textbf{1}, we  display the timeline with the key events for this complete classical evolution (from de Sitter to de Sitter) powered by a phenomenological creation of matter for different components. In the next two sections, we determine analytically the key results and the most interesting details of the timeline above. 
% Finally, to complete the thermodynamic description of cosmological fluids, it is important to determine the evolution of the temperature as the universe expands. The temperature law for ``adiabatic'' processes is well known in the literature and reads \cite{CL89,CLW92,LG92}
% \begin{equation}
%     \frac{\dot{T}}{T}=\left(\frac{\partial p}{\partial\rho}\right)_{n}\frac{\dot{n}}{n},
% \end{equation}
% which for the simple EoS (\ref{EoS}) has the straightforward solution $T\propto n^{w}$. In fact, in the standard cases where the particle concentration has a simple inverse relation with the commoving volume, i.e. $n\propto a^{-3}$, the temperature of radiation ($w=1/3$) scales as $T\propto a^{-1}$, while $T$ remains constant for matter ($w=0$), as expected.

%%%%%%%%%%%%%%%%%%%%%%%%%%%%%%%%%%%%%%%%%%%%%%%%%%%%%%%%%%%%%%%%%%%%%%%%%%%%
\section{A new complete Cosmology with particle creation} \label{NEW_CCDM}  
%%%%%%%%%%%%%%%%%%%%%%%%%%%%%%%%%%%%%%%%%%%%%%%%%%%%%%%%%%%%%%%%%%%%%%%%%%%%%%%%

Particle production rates should be obtained from quantum field theory (QFT) in curved space-time. However, the lack of a rigorous  theory in the presence of gravity has encouraged attempts in the phenomenological determination of $\Gamma_i$ over the past decades. For example, models powered by particle creation ($\Omega_{DE} \equiv0$) can mimic $\Lambda$CDM dynamics if particles are created as phenomenologically proposed in \cite{LJO10}. Complete cosmological scenarios with particle creation in which the evolution of the universe evolves smoothly  from an inflationary regime to a Friedmann-like universe that eventually accelerates have also been investigated \cite{LBC12,LGPB14}. These models are physically compelling because the two inflationary stages—from the earliest epoch to the present—are linked by a single underlying mechanism: the emergence of particles in curved spacetime.    

Since the standard cosmological model is also being  observationally challenged by two cosmological tensions ($H_0, S_8$), in what follows we investigate a multifluid model with matter creation. Let us suppose that the particle creation rates for radiation ($\Gamma_r$), dark matter ($\Gamma_{dm}$) and baryons ($\Gamma_{b}$) are, respectively:
%%%%%%%%%%%%%%%%%%%%%%%%%%%%%%%%%%%%%%%%%%%%%%%%%%%%%%%%%%%%%%%%%%%%%%%%%%%
\begin{align}
\Gamma_r&= 3H\beta_r + 3H\left(1-\beta_r\right) \left(\frac{H}{H_{I}} \right), \label{gammar} \\
\Gamma_{dm}&= 3H\beta_{dm} + 3H \alpha_{dm}\left(1-\beta_{dm}\right) \left(  \frac{\rho_{0}}{\rho_{dm}} \right), \label{gammadm} \\
\Gamma_{b}&= 3H\beta_{b} + 3H \alpha_{b}\left(1-\beta_{b}\right) \left(  \frac{\rho_{0}}{\rho_{b}} \right), \label{gammab}
\end{align}
%%%%%%%%%%%%%%%%%%%%%%%%%%%%%%%%%%%%%%%%%%%%%%%%%%%%%%%%%%%%%%%%%%%%%%%%%%%
where $\beta_i$ and $\alpha_i$ are dimensionless constants related to the production rate of each particle types ($i$) such as ($r$) radiation, ($dm$) dark matter, ($b$) baryons and etc, $\rho_{0}$ is the critical energy density (henceforth a subindex 0 denotes present day quantities). $H_I$ is the constant value of $H$ during the inflationary period, while $\rho_{dm}=n_{dm}m_{dm}$ and $\rho_b=n_bm_b$ are the rest energy density of dark matter particles and baryons, respectively. Note that for $\beta_r = 0$, the rate of creation ${\Gamma}_r$ reduces to the expression of \cite{LBC12}, while for $\beta_{dm}=\beta_b = 0$, the last two expressions above are the same defined in \cite{LJO10}. 

The Hubble parameter for each phase is easily obtained from equations (\ref{EE1}), (\ref{rhodot}) and the creation rates above, provided that $\omega_r=1/3$ for radiation and $\omega_m=0$ for dark matter and baryons. The density parameters are defined in the usual way,  $\Omega_r=\rho_{r0}/\rho_{0}$, $\Omega_{dm}=\rho_{dm0}/\rho_{0}$ and $\Omega_b=\rho_{b0}/\rho_{0}$ and from (\ref{EE1}) satisfy $\Omega_r + \Omega_{dm} + \Omega_b = 1$.

%\begin{figure}[!ht]
%\centering
%\includegraphics[scale=0.65]{timeline_alfa_beta_Atualizada.pdf}
%\caption{ Timeline  of the complete cosmological scenario from de initial Sitter ($H_I$)  to a final Sitter ($H_F$) stage (out of visual scale). Immediately after the unknown primordial quantum gravity state, ultra relativistic particles and radiation are created at the expenses of gravity (curvature) with the rate $\Gamma_I=3H_I$ and inflate the cosmos. Inflation ends ($\ddot a=0$) for $H=H_e$,  when the decelerating radiation phase begins ($\ddot a<0$). This radiation regime with creation ends at $z=z_{eq}$ marking the begin of the decelerating matter dominated phase with creation. At $z_{t}$, the creation pressure accelerates the cosmic expansion once more, which is observed today ($z=0$), until it finally induces a final de Sitter-like stage powered by $\Gamma_F=3H_F$ in the very distant future ($a\rightarrow\infty$). Figure adapted from \cite{PLS2024}.}
%\label{timeline}
%\end{figure}

\section{Cosmic Eras: From de Sitter (\texorpdfstring{$H_I$}{}) to de Sitter (\texorpdfstring{$H_F$}{})}

\subsection{From the primeval de Sitter-like inflation to a radiation regime. }\label{radiation}

At the very early stages of the universe, dark matter and baryons are negligible compared to radiation. 
%Although the evolution of all species of relativistic particles should be separately calculated, provided that each component might have a different production rate, we consider a single radiation fluid for simplicity. 
In this case, the differential equation governing the cosmic expansion rate of the early universe is determined by combining  (\ref{Hdot}) with the production rate (\ref{gammar}). We find:
\begin{equation}\label{Hubble2}
\dot{H} + 2 H^2 (1 - \beta_r )\left( 1 - \frac{H}{H_I}\right)=0,
\end{equation}
whose solution is easily obtained:
\begin{equation}\label{Hsolution}
H(a)=\frac{H_{I}}{1+Da^{2\left(1-\beta_r\right)}},
\end{equation}
where $D$ is a constant of integration. Given this solution, the evolution of the radiation energy density $\rho_r (a)$  follows directly from the Friedmann equation (\ref{EE1})
\begin{equation}\label{rhor(a)}
\rho_{r}(a)=\frac{\rho_{I}}{\left[1+Da^{2\left(1-\beta_r\right)}\right]^{2}}.
\end{equation}
Some interesting aspects on the early cosmic dynamics can be directly inferred from equations (\ref{Hubble2}) - (\ref{rhor(a)}).  For $\beta_r=0$, the dynamics of the primeval universe proposed in \cite{LBC12} is recovered from (\ref{Hubble2}). Moreover, regardless of the value of $\beta_r$, the above equation exhibits a de Sitter-like dynamics for $H=H_I$, since $(\dot{H}=0\rightarrow H=\textrm{constant})$ evolving smoothly to a decelerating radiation phase. Since both phases are dominated by radiation, clearly, there is no room to the so-called\, ``graceful" exit problem in these models. 

As expected, for $Da^{2(1-\beta_r)}\ll 1$, the above equations yield constant values for the Hubble scale and energy density, namely:  ($H,\rho$) $\equiv (H_I,\rho_I$). Conversely, for $Da^{2(1-\beta_r)}\gg1$, the Hubble parameter scales as  $H (a) \propto a^{-2(1-\beta_r)}$, and, similarly,  $\rho_{r}(a) \propto a^{-4(1-\beta_r)}$. The standard evolution for the radiation dominated universe is recovered for  $\beta_{r}=0$. However, for $\beta_r \neq 0$, these solutions suggest a change in the evolution of $\rho_{r}$ compared to the standard radiation dynamics. Thus, the continuous transition from a de Sitter-like regime to a radiation epoch is slightly modified by particle creation.

At this stage, a natural question arises: \textit{``What is the energy scale at which the Universe evolves from the primordial inflationary regime ($\ddot{a} > 0$) to the decelerating, radiation-dominated phase ($\ddot{a} < 0$)?"}

Inflation ends at the scale $H=H_e$, which can be easily evaluated from (\ref{Hubble2}) by calculating the deceleration parameter (see also \textbf{Figure 1}):    
\begin{equation}\label{qzrad} 
q=-\frac{\dot{H}}{H^{2}}-1= -1+2(1-\beta_r)\left(1-\frac{H}{H_{I}}\right).
\end{equation}
Hence, for $q \propto \ddot a =0$,  we see that the smooth transition defining the end of inflation occurs when the Hubble parameter reaches the value $H_e=H_I(1-2\beta_r)/(2-2\beta_r)$ 

% Furthermore, since the temperature law in ``adiabatic'' cases can be written as $T\propto n^{w}$, in the radiation dominated era ($w=1/3$) it becomes $T_r\propto n^{1/3}$. Additionally, it is easy to show from the local Gibbs Law, mentioned in section (\ref{PC}), that the radiation energy density in $\dot{\sigma}=0$ scenarios evolves as $\rho_r \propto n^{4/3}$. Thus, by contrasting $T_r\propto n^{1/3}$ with $\rho_r \propto n^{4/3}$, the well known relation $\rho_r \propto T_r^{4}$ is readily obtained. 
%Therefore, considering the energy density in equation (\ref{rhor(a)}), the temperature of radiation as a function of $a$ reads
% \begin{equation}\label{Tr(a)}
% T_r=T_{I}\left[1+Da^{2\left(1-\beta_r\right)}\right]^{-1/2}.
% \end{equation}
% The temperature in a de Sitter-like expansion remains constant, in agreement with the equation above for $Da^{2(1-\beta_r)}\ll1$ $(T_r\approx T_I)$. As the scale factor grows and reaches $Da^{2(1-\beta_r)}\gg 1$, $T_r$ behaves as $T_r\propto a^{(\beta_r -1)}$, which is different from the usual radiation scenario where the temperature evolves as $T_r\propto a^{-1}$. 

According to recent observations, the dynamics of the universe in alternative cosmological models should not be much different from $\Lambda$CDM, at both small and large cosmological scales \cite{Planck2018b,Brout2022}. Some investigations have also shown that cosmologies with mild distinctions from the standard model may shed some light on current cosmological problems, such as the tensions $H_0$ and $S_8$ \cite{valentino2021,Lucca2021}. 

In principle, a model in which the early radiation epoch is only minimally modified by $\beta_r \ll 1$ may help resolve or at least alleviate some of these difficulties, particularly if a small deviation is also introduced at low redshifts. Demonstrating this is the aim of the next subsection.

\subsection{From Matter to a Final de Sitter-like Phase}
%As discussed in the previous section, the universe transitioned naturally from a de Sitter-like
%phase to a radiation regime of expansion in the context of particle creation. Eventually,
In virtue of the universal expansion, the radiation energy density decreases, and finally, the material component (dark matter
and baryons) drives the cosmic dynamics. The evolution of the energy density for dark matter $\rho_{dm} (a)$ can be determined from EoS (\ref{EoS}), the continuity equation (\ref{rhodot}), and the particle production rate (\ref{gammadm}). Both radiation and matter impose a decelerating regime on cosmic dynamics. The rate of expansion H decreases and eventually becomes ($H\ll H_I$). 

Now, considering the particle creation rate (\ref{gammadm}) and the continuity equation for dark matter ($\omega=0$), we obtain
\begin{eqnarray}
    \frac{\dot{\rho}_{dm}}{\rho_{dm}}=-3\frac{\dot{a}}{a}(1-\beta_{dm})\left(1- \alpha_{dm}\frac{\rho_0}{\rho_{dm}}\right),
\end{eqnarray}
which can be rewritten as
\begin{eqnarray}
    \frac{d(\rho_{dm}-\alpha\rho_0)}{(\rho_{dm}-\alpha\rho_0)}= -3(1-\beta_{dm})\frac{da}{a}.
\end{eqnarray}
The solution to the equation above reads:
\begin{equation} \label{rhoma}
\rho_{dm} (a)= \left( \rho_{dm0} - \alpha_{dm} \rho_{0} \right) a^{-3(1-\beta_{dm})} + \alpha_{dm} \rho_{0}.
\end{equation}

Similarly, for baryons we take $p=0$ and combining  (\ref{rhodot}) with (\ref{gammab}), it follows that 
\begin{equation} \label{rhob}
\rho_{b} (a)= \left( \rho_{b0} - \alpha_{{b}} \rho_{0} \right) a^{-3(1-\beta_{b})} + \alpha_{b} \rho_{0}.
\end{equation}

%and analogous for the rest energy density of baryons ($\rho_{b}$), endowed
%with the production rate (\ref{gammab}) and $p=0$, whose solution is equivalent to equation (\ref{rhoma}), except with constants $\rho_{b0}$, $\alpha_{b}$ and $\beta_{b}$.
For simplicity, let's consider $\beta_{dm}\approx\beta_{b}$. By adding  (\ref{rhoma}) and (\ref{rhob}) we obtain:
\begin{equation} \label{rhombeta}
\rho_{m} (z)= \left(\rho_{m0} - \alpha \rho_{0} \right) \left(1+z\right)^{3(1-\beta)} + \alpha \rho_{0},
\end{equation}
where $\rho_m=\rho_{dm} + \rho_b$, $\alpha=\alpha_{dm} + \alpha_{b}$ and $\beta\equiv\beta_{dm}=\beta_{b}$. 

Note that whether $\alpha$ and $\beta$ are neglected, viz. $\alpha=\beta=0$, the standard case $n\propto\rho_{m}\propto a^{-3}$ is recovered. However, if $\beta$ is neglected but $\alpha \neq 0$, the solution describes the total dust energy density of the CCDM model \cite{LJO10}.

In the late phase of expansion,  dark matter and baryons dominate the energy density so that radiation can be neglected. In this scenario, the Hubble parameter $H$ is readily obtained from equation (\ref{rhombeta}), since deep in the matter dominated era $\rho_{m}\propto H^{2}$. In terms of the redshift it becomes:
\begin{equation} \label{Hlate}
H=H_0\left[  \left( 1 - \alpha \right) \left(1+z\right)^{3(1-\beta)} + \alpha \right]^{1/2},
\end{equation}
since matter is the dominant component at the late phase of the universe, i. e. $\Omega_m=\Omega_{dm}+\Omega_b=1$. A similar expression for $H(z)$ is obtained in Ref. \cite{jesus2017} in the context of dark matter creation with a conserved number of baryons (see also Ref. \cite{GCL2014}). 

It should be noted that galaxy clusters and other observations suggest $\Omega_m{\approx0.3}$. However, as argued in the LJO paper \cite{LJO10}, the low redshift constraints on $\Omega_m$ in models with particle creation describe an effective matter energy density, $\Omega_{meff}=\Omega_m -\alpha\approx0.3$, which is responsible for the portion of matter that agglomerates in clusters. In this perspective, the remnant fraction of dust, that is $1-\Omega_{m}\approx 0.7$, is presumed to be at the background level as homogeneously distributed throughout the universe. The case for a first order perturbative level will be discussed in a forthcoming communication. 
% This solution is comparable to the expansion rate of the model proposed by Jesus and Collaborators \cite{GCL2014,jesus2017}.

Eventually, as the redshift decreases and approaches the value $z=-1$ ($a\rightarrow \infty$), $H$ becomes approximately constant $H_F=H(a\rightarrow\infty) \approx H_0\sqrt{\alpha}$. A constant $H_{F}=\dot{a}/a$ implies a de Sitter-like expansion ($a\propto e^{H_{F} t}$) at the very late stage of the universe, without dark energy, coinciding with the CCDM model (see again \textbf{Figure  1}).

\begin{figure}
    \centering
    \includegraphics[width=1.0\linewidth]{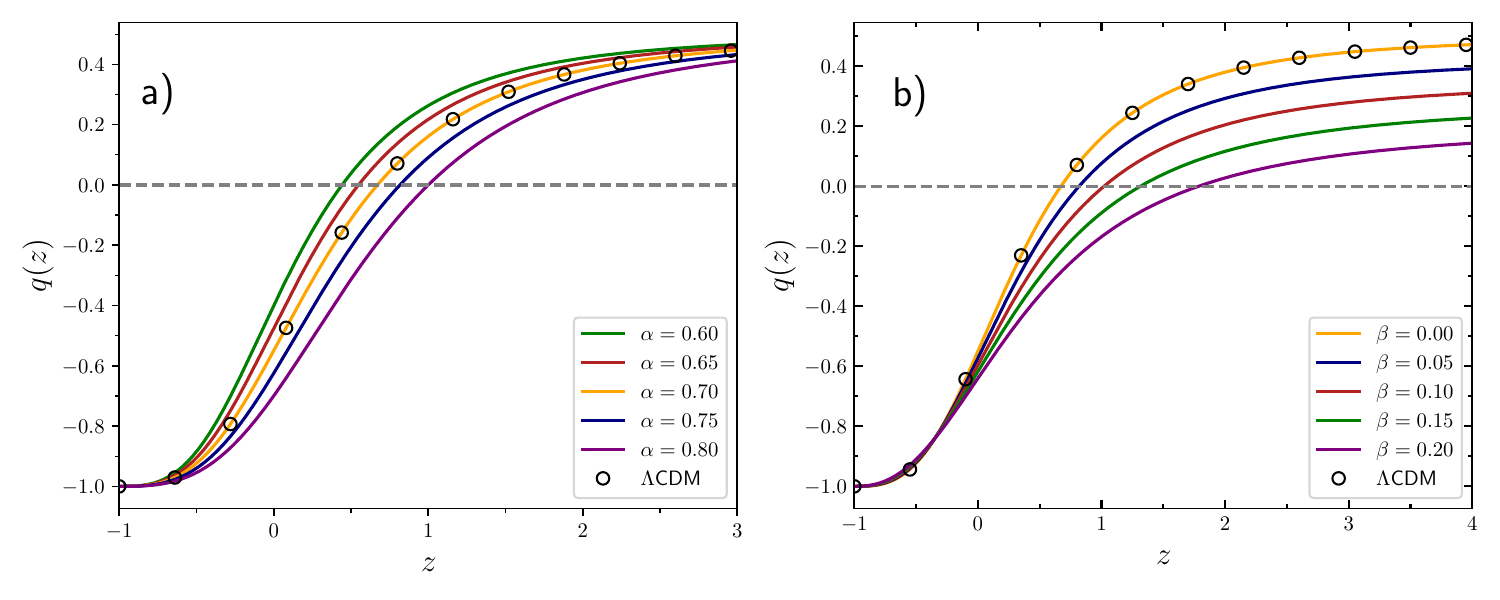}
    \caption{The evolution of the deceleration parameter (\ref{qzab}) with respect to the redshift $z$. Figure a) depicts the evolution of $q(z)$ for different values of $\alpha$ with $\beta=0$. Figure b) represents the changes in the evolution of $q(z)$ caused by the contributions of different values of $\beta$ considering $\alpha=0.7$ (value predicted by the CCDM model). In both plots the predictions of the $\Lambda$CDM model (empty black circles) coincides with the CCDM model (yellow lines, $\alpha=0.7,\beta=0$).}
    \label{figqzab}
\end{figure}

In the matter dominated phase, it is also interesting to deduce the deceleration parameter, in particular, the expression for the transition redshift ($z_t$). Both can be calculated by combining the definition of $q$ with  
 the Hubble parameter (\ref{Hlate}). It reads:
\begin{equation}\label{qzab}
    q=\frac{3\left(1-\beta\right)}{2}\left[\frac{\left(1-\alpha\right)\left(1+z\right)^{3\left(1-\beta\right)}}{\left(1-\alpha\right)\left(1+z\right)^{3\left(1-\beta\right)}+\alpha}\right]-1.
\end{equation} 
Hence, as $z\rightarrow-1$, the deceleration parameter approaches $q=-1$, in agreement with the de Sitter-like dynamics in the last stage of expansion. The transition from the decelerating regime to the final accelerating phase of the cosmic expansion occurs at the redshift transition, defined again for $q=0$. Hence, from equation (\ref{qzab}), $z_t$ can be written as
\begin{equation}\label{zt_ab}
    z_{t}=\left[\frac{2\alpha}{\left(1-\alpha\right)\left(1-3\beta\right)}\right]^{1/3\left(1-\beta\right)}-1.
\end{equation}
Note that $z_t$ diverges for $\beta = 1/3$, thus suggesting that $\beta\leq 1/3$. Observe also that this transition occurs at higher or lower redshifts, when compared to the CCDM case ($\beta=0$), depending on the values of $\alpha$ and $\beta$ (see Figure \ref{figqzab}).

\section{Matter Phase: A Comparison with Observations}\label{analysis}

The free parameters of the cosmological model proposed in this paper are constrained with observational data of apparent magnitudes from SNe Ia combined with Cepheid distances sampled by the Pantheon+SH0ES collaboration \cite{Brout2022}, along with $H(z)$ data from cosmic chronometers \cite{MorescoEtAl20,MorescoEtAl22}.

\subsection{Data and Methodology}

Measurements of apparent magnitudes from type Ia Supernovae (SNe Ia) and Cepheid distances are independent of cosmological model choice, since it depends only on the astrophysical properties of these objects. Consequently, these measurements are suitable for constraining cosmic parameters to observations. The correspondence between theory and experiment for SNe Ia can be obtained from the equation 
\begin{equation}
    m(\Theta)= M + 5\log_{10}(d_L) + 25,
\end{equation}
where $m$ and $M$ are respectively the apparent and absolute magnitudes of Supernovae, $\Theta=(M,H_o, \alpha, \beta)$ represents the free cosmological parameters of the model and $d_L$ is the luminosity distance 
\begin{equation}
    d_L = c(1+z)\int_{0}^{z} \frac{dz^{\prime}}{ H(z^{\prime};\Theta)}
\end{equation}
where c is the speed of light. 

Currently, one of the largest SNe Ia dataset available is the Pantheon+SH0ES compilation \cite{Brout2022}, containing 1701 light curves from 1550 distinct SNe Ia with redshifts from $0.001<z<2.26$, combined with Cepheid distances from SH0ES. The Cepheid distance calibration contributes significantly to the estimates of the free parameters ($\Theta$) by imposing strong constraints on $H_0$. 

In addition to Supernovae, we use $H(z)$ data from cosmic chronometers (CC) to strengthen the constraints on the parameters $\Theta$. This observation is also independent of the cosmological model choice, since it depends only on the age of astronomical objects determined by their chemical evolution. In this case, $H(z)$ can be obtained from the simple relation $H(z)=-[(1+z)dt/dz]^{-1}$ \cite{JimenezLoeb02}. For this test, we consider 32 CC $H(z)$ data with systematic errors from Moresco et al. \cite{MorescoEtAl20,MorescoEtAl22}. 

Assuming the Bayes' Theorem, given an observational dataset $D$, we evaluate the posterior $p(\theta|D)\propto\pi(\theta)\mathcal{L}(\theta|D)$ with parameters $\theta=(M,H_0,\alpha,b)$, in which $\pi$ is the flat prior and $\mathcal{L}=\mathcal{L_{SN}}\mathcal{L_{H}}$ is the likelihood of the combined datasets with $\mathcal{L_{SN}}$ regarding the Pantheon+SH0ES compilation and $\mathcal{L_{H}}$ the $H(z)$ data from CC. These likelihoods can be written in the form $\mathcal{L}_{(SN,H)}\propto e^{-\chi_{(SN,H)}^2/2}$ where
\begin{equation}
    \chi_{(SN,H)}^2=\sum_{i,j}\left[y(z_i,\theta)-y_{obs,i}\right]C_{ij}^{-1}\left[y(z_j,\theta)-y_{obs,j}\right],
\end{equation}
is the Chi-squared distribution, $C_{ij}$ is the covariance matrix and $y$ represent the theoretical $y(z,\Theta)$ and observational $y_{obs}$ values of the apparent magnitude $m$ for the SNe Ia and $H$ for the CC $H(z)$ datasets. The theoretical prediction of the Hubble parameter used in these analysis is the one in equation (\ref{Hlate}) due to the low redshifts of the observations. The posteriors are tested by generating Monte Carlo-Markov Chains (MCMC) and sampling $p(\theta|D)$ using the emcee python package \cite{ForemanMackey13}. We also used the getdist python package to generate smooth MCMC results \cite{getdist}.

\subsection{Basic Results}

\begin{figure}[ht]
    \centering
    \includegraphics[width=1\textwidth]{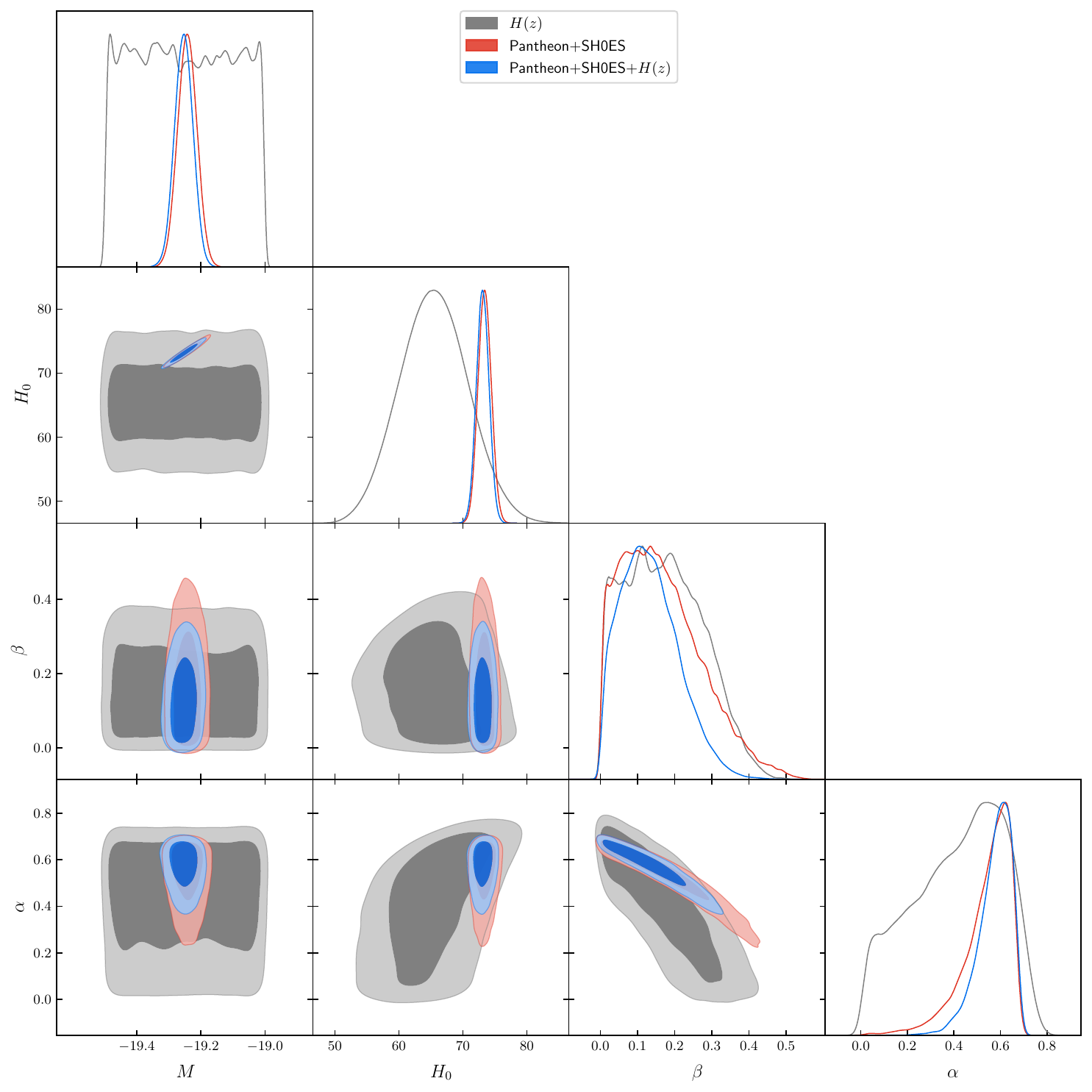}
    \caption{ Constraints on $M$, $H_0$, $\alpha$, $\beta$, using SNe Ia observations and Cepheid calibrations from the Pantheon+SH0ES compilation \cite{Brout2022}, along with $H(z)$ data from cosmic chronometers \cite{MorescoEtAl20,MorescoEtAl22}.}
    \label{triangle1}
\end{figure}

The constraints on the free parameters of the proposed model $(M,H_o,\alpha,\beta)$ are displayed in Figure \ref{triangle1}, for the 1$\sigma$ and 2$\sigma$ confidence levels (c.l.), regarding all the datasets discussed in the previous section. Note in Figure (\ref{triangle1}) that the Pantheon+SH0ES compilation constraints are more limiting than the cosmic chronometers (CC) results. Nonetheless, the combination of SNe Ia data with CC observations $H(z)$ yields stronger constraints. 

The best fit for the free parameters can be visualized in Table (\ref{tab1}).
The results $\alpha\approx0.57$ and $\beta\approx 0.13$ implies that the provided data suggests a cosmic dynamics slightly different from the CCDM and the standard scenarios (see Figure \ref{figqzab}). The moment of exchange from the inflationary period to a modified radiation phase can be inferred from the results above in terms of $H_I$, as $H_e=H_I(1-2\beta)/(2-2\beta)\approx0.42 H_I$.  These results leads to an effective energy density of matter $\Omega_{meff}=1-\alpha\approx 0.42$, which is bigger than the standard value $\Omega_{meff}(\text{CCDM}/\Lambda \text{CDM})\approx 0.3$. 
%\ref{fig_at}, \ref{tz_best} and
%With regard to the redshift of equivalence, a proper estimate of $z_{eq}$ requires stronger constraints on $\beta$, since the results in Table (\ref{tab1}) yields values from $10^{3}$ up to $10^{4}$, as discussed in section (\ref{radiation}).
\begin{table}[ht] 
    \centering
   \begin{tabular} { l c c}
   
\hline
 Parameter & \ 68\% C.L & 95\% C.L.\\
\hline
{\boldmath$M              $} & $-19.253^{+0.028}_{-0.028}   $ & $-19.254^{+0.056}_{-0.056} $\\
{\boldmath$H_0            $} & $73.1^{+0.98}_{-0.98}        $ & $73.1^{+2.0}_{-1.9}        $\\
{\boldmath$\alpha         $} & $0.576^{+0.088}_{-0.040}     $ & $0.58^{+0.12}_{-0.15}   $\\
{\boldmath$ \beta              $} & $0.133^{+0.057}_{-0.098}                     $ & $ 0.13^{+0.15}_{-0.13}                   $\\
\hline
\end{tabular}
    \caption{ Best fit of the free parameters $M$, $H_o$, $\alpha$ and $\beta$ at 68\% and 95\% confidence levels (C.L.) from the combined datasets Pantheon+SH0ES \cite{Brout2022} and $H(z)$ from cosmic chronometers \cite{MorescoEtAl20,MorescoEtAl22}.}
    \label{tab1}
\end{table}

\begin{figure}[!ht]
\centering
\includegraphics[scale=0.7]{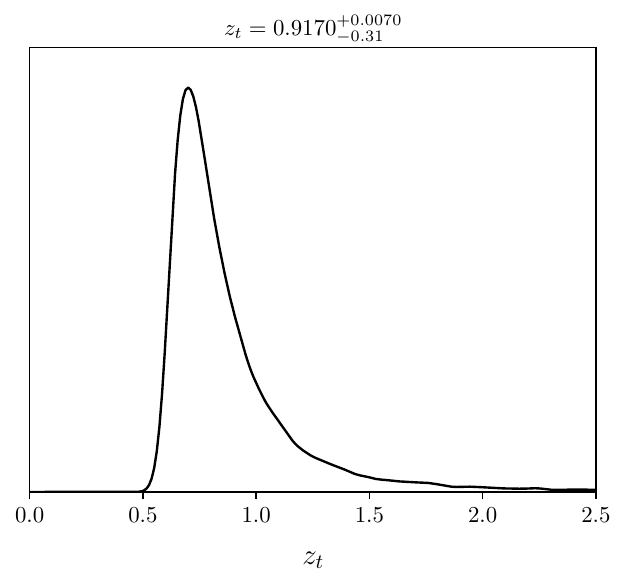}
\caption{ The Likelihood of the transition redshift $z_t$, regarding the constraints on the parameters $\alpha$ and $\beta$ from the compilation Pantheon+SH0ES \cite{Brout2022} with $H(z)$ CC data \cite{MorescoEtAl20,MorescoEtAl22} and the theoretical prediction for $z_t$ in equation (\ref{zt_ab}), which constrains $z_t$ to $z_t=0.917^{+0.0070+0.67}_{-0.31-0.38}$ at 1$\sigma$ and $2\sigma$ c.l. }
\label{zt_best}
\end{figure}

\begin{figure}[ht]
    \centering
    \includegraphics[width=1\textwidth]{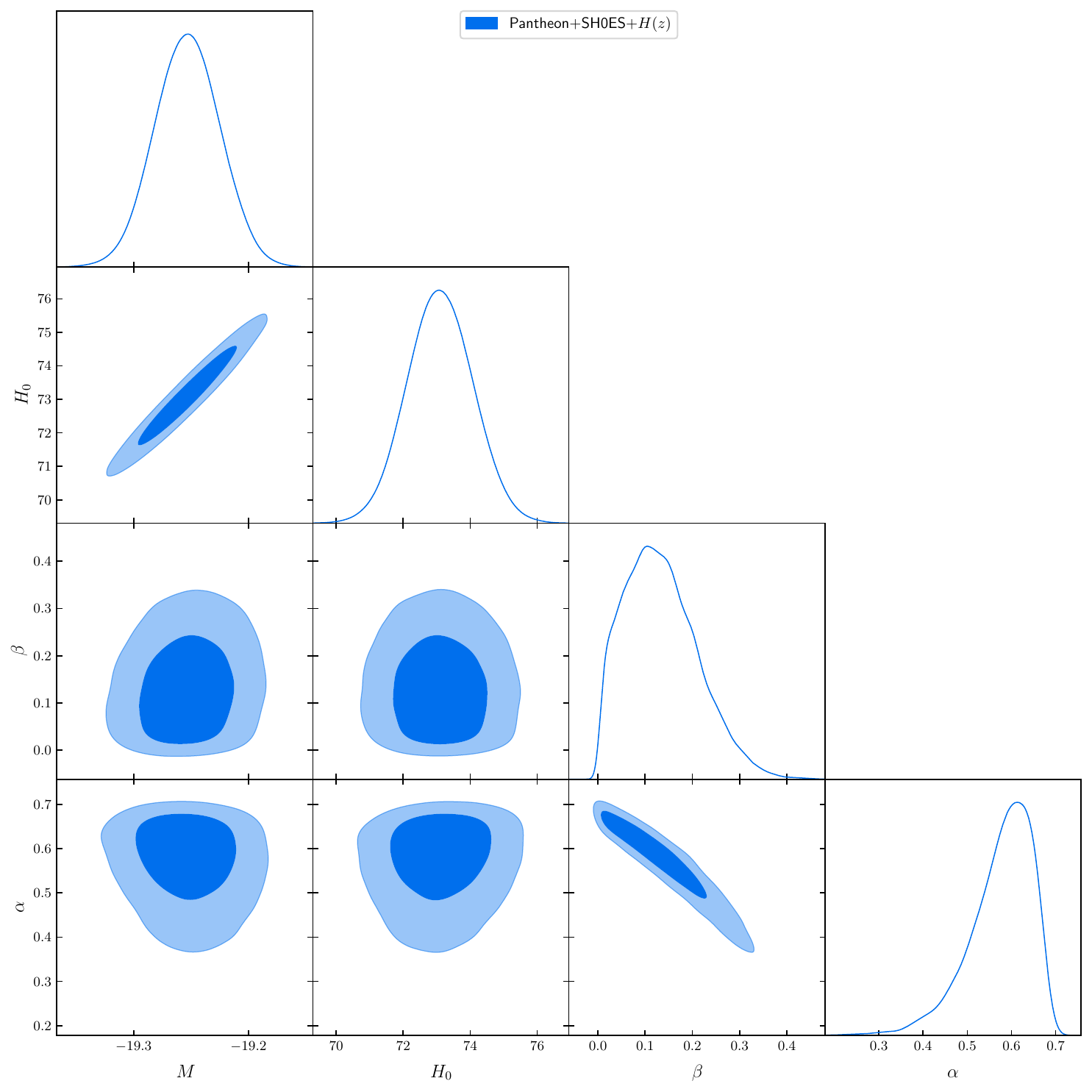}
    \caption{Combined constraints on $M$, $H_0$, $\alpha$, $\beta$, using the datasets Pantheon+SH0ES \cite{Brout2022} and $H(z)$ from cosmic chronometers (CC) \cite{MorescoEtAl20,MorescoEtAl22}.}
    \label{triangle2}
\end{figure}

The constraints on $\alpha$ and $\beta$ can also be used to construct the probability distribution function (PDF) for the transition redshift $z_t$ in the equation (\ref{zt_ab}). It is important to recall, however, that $z_t$ exhibit problems for $\beta \geq 1/3$ in this model (see equation \ref{zt_ab}). Thus, setting the limits on the flat prior of $\beta$ at $\beta\in [0, 0.33]$, the PDF of $z_t$ can be visualized in Figure (\ref{zt_best}), where $z_t=0.917^{+0.007}_{-0.31}$ at 1$\sigma$ c.l. This result suggests a slightly higher $z_t$ than the standard model $z_t(\Lambda\textrm{CDM/CCDM})\approx0.71$.

% The age of the universe according to this model can be calculated from equation (\ref{t}) with the best fit for the parameters in Table (\ref{tab1}), and the result is $t_o\approx H_o^{-1}\left[1 + \mathcal{O}(10^{-3})\right]$, which is equivalent to $t_o\approx 13.3$ billion years using $H_o=73.1$ Km$\cdot$s$^{-1}$$\cdot$Mpc$^{-1}$ from Table (\ref{tab1}).

Finally, the value of $H_F=H_0\sqrt{\alpha}$ can also be evaluated with the outcome from the analysis above, and it yields $H_F\approx 0.75H_0$ or $H_F\approx 55.2$ Km$\cdot$s$^{-1}$$\cdot$Mpc$^{-1}$ using the constraints on $H_0$ from Table (\ref{tab1}).

Considering that this analysis was computed at the background level of cosmology, the parameters are well constrained by the combined data. Surely, more background observations along with probes in the perturbed level of cosmology can provide stronger constraints on the free parameter of the model. It is also important to mention that strong correlations between the parameters $M$ and $H_0$, as well as between $\alpha$ and $\beta$, can be seen in Fig.(\ref{triangle2}). The former have already been discussed in reference \cite{Brout2022}, thus it will not be exploited in this paper. On the other hand, a brief discussion regarding the correlation between $\alpha$ and $\beta$ is addressed in the Appendix (\ref{xy}). 

%%%%%%%%%%%%%%%%%%%%%%%%%%%%%%%%%%%%%%%%%%%%%%%%%%%%%%%%%%%%%%%%%%%%%%%%%%%%%%%%%%%%%%%%%%%
\newpage
\section{Final Remarks}

In this work, we proposed a coherent classical and singularity-free cosmology that removes the need for dark energy. The model describes a smooth evolution from an initial de Sitter–like inflationary regime to a Friedmann phase modified by gravitationally induced particle creation, culminating in a final de Sitter–like era associated with the current cosmic acceleration. The negative pressure generated by the adiabatic particle-creation process drives both accelerated stages without resorting to dark energy, exotic fluids, extra dimensions, or modifications of general relativity. As a consequence, several persistent issues of the $\Lambda$CDM framework—most notably the cosmological-constant problem and the coincidence puzzle—are absent. In addition, because inflation arises naturally within the radiation era, the traditional horizon, flatness  and the ``graceful-exit"  problems are automatically resolved.

%In this paper,  a new and complete classical cosmology, without dark energy, has been proposed. In this scenario, the universe smoothly transits from an early de Sitter-like stage (Inflation) to a Friedmann-like cosmology modified by particle creation that eventually enters a final de Sitter-like phase (current accelerated expansion) of the cosmic evolution. Due to the negative pressure that naturally emerges from the ``adiabatic'' particle creation mechanism, this model provides an explanation for the two accelerated expansions of the universe without evoking dark energy, other exotic fluids, extra-dimensions or modified gravity. Since dark energy is absent in this scenario, some of the chronic problems that plague the $\Lambda$CDM model are inexistent in the proposed model such as the cosmological constant problem and the coincidence mystery. Moreover, since inflation is a natural mechanism in this scenario, the horizon and flatness problems are readily solved.

The model provides a cosmological description with simple analytical solutions that mildly departures from the late-time $\Lambda$CDM dynamics, providing corrections of $\beta$ at high and low redshifts. After the unknown quantum-gravity regime (see Figure 1), the classical model naturally transits from an unstable de Sitter expansion, at $H=H_I$, to a decelerated radiation phase modified by the gravitationally induced particle creation. After $z_{eq}$, the energy density of radiation becomes less significant than dust, and a matter dominated phase changed by $\beta$ and $\alpha$ begins. As the universe expands, the negative pressure connected with the mechanism of particle creation induces the cosmic accelerated expansion observed today. Eventually, as the matter energy density decreases, the universe smoothly enters a final and stable de Sitter-like expansion ($H=H_F$) due to the effects of particle creation.
%Furthermore, it was shown that the temperature law is also modified by the contributions of $\beta$ in the primeval cosmic scenario.

The cosmological parameters of the model were constrained by type Ia SNe observations with Cepheid calibrations from the Pantheon+SH0ES compilation along with $H(z)$ data from cosmic chronometers (CC).  
The SNe Ia dataset imposes stronger constraints on the cosmological parameters, although the $H(z)$ data also contribute to limit the parameters. The best fit $\beta=0.133_{-0.098}^{+0.057}$, at $68\%$ confidence level, indicates a preference of the observations for a mildly distinct dynamics compared to the standard model. In addition, the result $\alpha=0.576^{+0.088}_{-0.040}$ suggests a higher value of $\Omega_{meff}\approx 0.42$ compared to the  CCDM and $\Lambda$CDM models with $\Omega_{meff}(\Lambda\text{CDM/CCDM})\approx 0.3$, which must be investigated in the context of galaxy clusters formation. These results also indicate  that the transition from the decelerated to the current accelerated phase of cosmic expansion occurred slightly sooner than the standard scenarios, at $z_t\approx0.9$ as opposed to $z_{t}(\Lambda\text{CDM/CCDM})\approx0.71$. This effect is expected, since $\beta$ increases the negative pressure of particle creation responsible for accelerating the universe

Finally, we emphasize that the mildly dynamical deviations from the standard $\Lambda$CDM scenario introduced by the parameters $\alpha$ and $\beta$ may offer a natural explanation for the current $H_0$ and $S_8$ tensions. Additional background probes, together with perturbation-level cosmological observations, should be confronted with the proposed model in order to obtain tighter and more robust constraints on its parameters. This extended analysis is currently underway and will be presented in a forthcoming paper.

%%%%%%%%%%-------------------------------
%%%%%%%%%---------------------------------

\section{Appendix}

\subsection{Equivalence Redshift}

The radiation energy density evolves as $\rho_{r}\propto(1+z)^{4(1-\beta_r)}$ when $D(1+z)^{-2(1-\beta_r)}\gg1$, as discussed in section (\ref{radiation}). On the other hand, the constant terms in the energy density of dark matter and baryons can be neglected at high redshifts. In the epoch of equivalence between the radiation and matter energy densities, at the equivalence redshift $z_{eq}$, the relation
\begin{equation}
    \Omega_{r}\left(1+z_{eq}\right)^{4(1-\beta_r)}=\left(\Omega_{m}-\alpha\right)\left(1+z_{eq}\right)^{3\left(1-\beta\right)},
\end{equation}
must be satisfied. In addition, the sum of the energy density fractions of all components in a flat universe, including radiation, is equal to unity: $\Omega_r+\Omega_m=1$. Thus, isolating $z_{eq}$ in the equation above yields
\begin{equation}\label{zeq}
z_{eq}=\left(\frac{1-\alpha-\Omega_{r}}{\Omega_{r}}\right)^{1/\left(1-4\beta_r +3\beta\right)}-1.
\end{equation}
The equivalence redshift above is sensitive to variations on $\beta_{r}$ and $\beta$, hence an estimate of $z_{eq}$ requires strong constraints on these parameters. Section (\ref{analysis}) provides constraints on $\alpha$ and $\beta$ from low redshift observations. Nonetheless, astronomical observations of high redshift events, such as the anisotropies of the cosmic microwave background radiation, should be used to constraint $\beta_r$ and possibly $z_{eq}$ in forthcoming papers.

\subsection{Correlation Coefficient Reduction}\label{xy}

A pronounced degeneracy between $\alpha$ and $\beta$ is evident in Fig.5, reflecting their strong mutual correlation. Indeed, the correlation coefficient between these parameters is $r_{\alpha\beta} = -0.9475$ [65]. Such a high degree of correlation hampers the independent determination of $\alpha$ and $\beta$.

%A degeneracy between $\alpha$ and $\beta$ can be seen in Figure (\ref{triangle2}), due to their strong correlation. In fact, the correlation coefficient between these two parameters is $r_{\alpha\beta}=-0.9475$ \cite{correlation}. This degeneracy compromise the independent determination of $\alpha$ and $\beta$.

\begin{figure}[ht]
    \centering
    \includegraphics[width=0.5\textwidth]{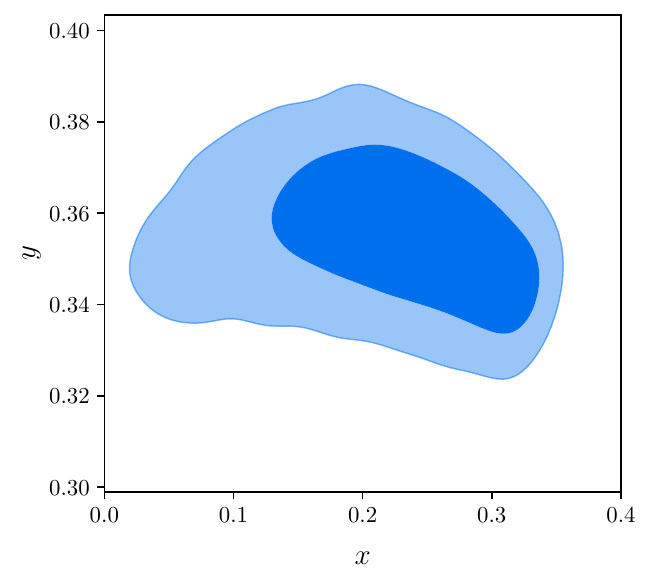}
    \caption{Constraints on $x=\alpha-\beta$ and $y=\alpha+\beta$ for the combined Pantheon+SH0ES \cite{Brout2022} and $H(z)$ CC \cite{MorescoEtAl20,MorescoEtAl22} datasets.}
    \label{figxy}
\end{figure}

In an attempt to diminish this degeneracy, the parametrization $x=\alpha-\beta$ and $y=\alpha+\beta$ can be used, as proposed in the literature \cite{jesus2017}. The constraints on the new set of parameters yield $x=0.22^{+0.12}_{-0.15}$ and $y=0.355^{+0.025}_{-0.024}$. Indeed, the degeneracy is suppressed in the $(x,y)$ plane (see Figure \ref{figxy}), and a considerably smaller correlation coefficient $r_{xy}=-0.1685$ is obtained with regards to $x$ and $y$.

\begin{acknowledgments}
PWRL is financially supported by the National Council for Scientific and Technological Development (CNPq) under grant 140100/2021-0. JASL is also partially supported by the Brazilian agencies, CNPq under Grant 310038/2019-7, CAPES
(grant 88881.068485/2014) and FAPESP (LLAMA Project 11/51676-9). 
\end{acknowledgments}

%%%%%%%%%%%%%%%%%%%%%%%%%%%%%%%%%%%%%%%%%%%%%%%%%%%%%%%%%%%%%%%%%%%%%%%%%%%%%%%%%%%%%%%%%%%%%%%%%%%%
%\appendix

\bigskip

%%%%%%%%%%%%%%%%%%%%%%%%%%%%%%%%%%%%%%%%%%%%%%%%%%%%%%%%%%%%%%%%%%%%%%%%%%%%%%%%%%%%%%%%%%%%%%%%%%%%

\end{document}